# Direct observation of ballistic Brownian motion on a single particle

Rongxin Huang,[1] Branimir Lukic[2], Sylvia Jeney,[2] Ernst-Ludwig Florin[1]*

[1]Center for Nonlinear Dynamics, University of Texas, Austin, Texas 78712, USA
[2]Institut de Physique de la Matière Complexe, Ecole Polytechnique Fédérale de Lausanne (EPFL), CH-1015 Lausanne, Switzerland

*To whom correspondence should be addressed. E-mail: florin@chaos.utexas.edu

**Abstract**

At fast timescales, the self-similarity of random Brownian motion is expected to break down and be replaced by ballistic motion. So far, an experimental verification of this prediction has been out of reach due to a lack of instrumentation fast and precise enough to capture this motion. With a newly developed detector, we have been able to observe the Brownian motion of a single particle in an optical trap with 75 MHz bandwidth and sub-Ångstrom spatial precision. We report the first measurements of ballistic Brownian motion as well as the first determination of the velocity autocorrelation function of a Brownian particle. The data are in excellent agreement with theoretical predictions taking into account the inertia of the particle and the surrounding fluid as well as hydrodynamic memory effects.

**Introduction**

The botanist Robert Brown was the first to systematically investigate the erratic motion of suspended microscopic particles (1). Under Brown's microscope, each step seemed completely independent of the previous step. It is this randomness of motion that is the hallmark of Brownian motion. The origin of this effect was first successfully explained by Einstein as the amplification of the statistical fluctuations of the surrounding fluid molecules (2). Since then, the theory of Brownian motion has found broad applications in the description of phenomena in many fields in science (3) and even in financial models (4). Stochastic changes are the common thread that connects this wide range of phenomena.

The randomness of the transport process in Brownian motion has specific consequences. The average position of the particle remains unchanged while the mean square displacement (MSD) that describes the average distance explored by the particle increases linearly in time: MSD(t)=2Dt, where D is the diffusion constant and t the time. In contrast, the mean square displacement for directed transport with constant speed increases with the square of time. Due to the fractal nature of a Brownian particle's trajectory, its velocity is poorly defined. We do not know the length of the path traveled in a time interval and hence not the velocity. Einstein was aware that his statistical approach must break down at short times, because the particle's inertia must become significant and thereby lead to a correlated motion (5). The first mathematical description of a particle's dynamics over the entire time domain was provided by Langevin (6). The particle's inertia results in an exponentially damped velocity autocorrelation function with a characteristic time scale $\tau_p = m/\gamma$, where $m$ is the mass of a spherical particle and $\gamma$ is the Stokes' viscous drag coefficient. The velocity autocorrelation time is about 100 nanoseconds for a 1 micrometer polystyrene particle in water at room temperature. The mean-square displacement (MSD) approaches $2Dt$ at a long time scale, as expected for random Brownian motion, but approaches $k_B T/m \cdot t^2$ below $\tau_p$, where $k_B$ is the Boltzmann constant and $T$ is the temperature. In a simplified picture, the $t^2$ dependence indicates ballistic motion of the particle, that is, the particle flies in a straight line with constant speed before collisions with the water molecules slow it down and randomize its direction. Langevin's theory successfully took the particle's inertia into account, but he neglected another important factor: the inertia of the surrounding fluid. When the suspended particle moves, it has to drag and displace the surrounding fluid which introduces two effects: First, the co-moving fluid leads to an increased effective mass $m^*$ which is the sum of the mass of the particle and half the mass of the displaced fluid. Second, the surrounding fluid develops flow patterns which are long-lived and influence the particle's dynamics at later times--the so-called hydrodynamic memory effect (7, 8). The fluid inertia introduces a new characteristic time scale $\tau_f = r^2 \rho_f/\eta_f$, which is the time it takes for a vortex to diffuse over a distance of one particle radius $r$ in a fluid of density $\rho_f$ and viscosity $\eta_f$. $\tau_f$ and $\tau_p$ can be related by $\tau_f/\tau_p = 2\rho_p/9\rho_f$. Theories that include the fluid inertia effects were first developed in 1945 (9), and later brought to wide attention by Hinch in 1975 (10). The MSD derived by Hinch is rather complex and, therefore, we consider only three distinct cases. At very long time scales, the diffusive regime, all inertia effects die out and the motion is completely random as assumed in Einstein's theory of Brownian motion. At the time scale below $\tau_p$, the motion becomes ballistic and is dominated by the inertia of the particle and its surrounding fluid. The MSD in the ballistic regime approaches $k_B T/m^* \cdot t^2$. Between the ballistic and the diffusive regime lies a transitional non-diffusive regime, where motion is dominated by the hydrodynamic memory effect.

There have been several attempts to measure the transition from ballistic to random Brownian motion, but the approaches were limited by insufficient bandwidth, spatial resolution or by multiple particle interactions (11, 12, 13). Recently, non-diffusive Brownian motion of single particles has been observed on the time scale of $\tau_f$ and the dominance of the hydrodynamic memory effect over the particle's inertia has been shown (13). However, Brownian motion of a single particle in the ballistic regime where the motion is dominated by the inertia of the particle and its surrounding fluid has never been observed and thus theories for this regime have never been tested.

The main difficulty in the direct observation of ballistic Brownian motion of a single particle is the coupling between length scales and time scales. For instance, a 1 μm particle will on average move about 1 nm within 1 μs. Therefore, resolving Brownian motion at the ballistic regime requires a combination of high temporal and spatial resolution of the position detector. Although progress has been made in achieving higher spatial resolution, temporal resolution is still lacking and so far, no instrument has met both requirements simultaneously.

With the recent improvement of position detector bandwidth in our photonic force microscope (14), we are now able to directly observe Brownian motion of single particles confined by an optical trap with a bandwidth of 75 MHz and for the first time verify the existence of ballistic Brownian Motion. The optical trap helps to confine the particle in the observation volume where its displacement is measured by laser interferometry (15). The detector measures the displacement of the particle along one lateral dimension. The confinement by the optical trap introduces another time scale $\tau_k = \gamma/K$, where $K$ is the stiffness of the optical trapping potential, which is typically orders of magnitude larger than the time scale for ballistic motion. As long as $\tau_k$ is much longer than $\tau_f$, the motion can be considered free for time scales shorter than $\tau_k$ (13). In our experiment, $\tau_k$ is typically two orders of magnitude larger than $\tau_f$.

A schematic representation of our experimental setup is shown in Figure 1. The sample chamber is filled with particles at a concentration of about 1 particle per 200 x 200 x 200 μm$^3$ volume. This low concentration avoids hydrodynamic interaction between particles and trapping of multiple particles during the experiment. The trapped particle is kept away from the bottom coverslip with a distance at least 10 times the particle's diameter, ensuring that the hydrodynamic interaction between the particle and the bottom coverslip is negligible. While the optically trapped particle undergoes Brownian motion, its position time trace is recorded with a high bandwidth detector system. The detector output (75 MHz bandwidth) is recorded with a high bandwidth digitizer at 100 MHz for 40 ms. The length of the data segment is limited by the size of the digitizer's on-board memory. For calculating the velocity autocorrelation function, multiple data segments on the same particle have been recorded for averaging.

**Results**

Figure 2 shows the MSD of a 1 μm silica particle calculated over five orders of magnitude in timescales from about 10 ns to 1 ms. The MSD initially increases, and then reaches a plateau around 0.1 ms. The plateau indicates that the particle is confined by the optical trap. For shorter times the particle undergoes free, but not necessarily random Brownian motion. To obtain the position calibration, the MSD was fitted in the intermediate regime (the regime not yet affected by the trapping potential) with the Hinch theory for a free Brownian particle with the particle radius and a multiplication factor taken as fitting parameters (see Supporting Material). The temporal resolution is defined as the time where the signal-to-noise-ratio reaches unity. For a 1 μm silica particle, our temporal resolution is about 40 ns. At this temporal resolution, we can resolve a MSD as small as 0.0035 nm$^2$, corresponding to a sub-Angstrom spatial resolution, and smaller than the size of a hydrogen atom. The temporal resolution of the setup is not limited by the detector's theoretical bandwidth, but by background noise, including the laser noise, as well as electronic noise from the pre-amplifier and the digitizer. Mechanical perturbations of the setup are typically below 1 KHz, which is three orders of magnitude slower than the time scale of interest. We have performed experiments on silica (density =1.96g/ml), polystyrene (density=1.05g/ml) and resin (density=1.5g/ml) particles with sizes ranging from 500 nm to 3.5 μm. The temporal resolution varies but it is always smaller than the characteristic time scale of the particles' inertia effect $\tau_p$. At this temporal resolution, the particle's inertia effect on Brownian motion is expected to be clearly visible.

Figure 3 (a) shows the normalized MSD for silica and polystyrene particles of the same size. The MSD has been normalized to the behavior seen in the random diffusive regime, *2Dt*, and the time axis has been normalized to $\tau_f$. The fact that the normalized MSD of both particles deviates from unity implies that they do not undergo random motions at the time scales shown. The two curves overlap at longer times and separate more as they approach the shorter time regime. Since both silica and polystyrene particles have the same diameter, they must displace the same amount of fluid, and the fluid inertia effect is the same. Therefore, their digression is an indication of the effect of only the particle's inertia on its motion. However, silica has a higher density than polystyrene, and so will move more slowly after receiving an impulse from the surrounding water molecules. The difference in the MSD values is discernible right around $\tau_f$, and becomes manifest at shorter time scales.   At the resolution limit of 0.07 $\tau_f$, the difference reaches a maximum value of 100%.   At times above 2 $\tau_f$, the two curves overlap, indicating that the motion in this temporal regime is completely dominated by the fluid inertia.   The ballistic motion in which the particle's motion is dominated by its own inertia is clearly resolved in the experiments.

Figure 3 (b) shows the MSD normalized to the behavior in the ballistic regime $k_BT/m^* \bullet t^2$ for silica particles with different sizes.   For a 1 μm silica particle, $\tau_p$ is very close to the temporal resolution, but for a 2.5 μm silica particle, the temporal resolution is far smaller than $\tau_p$.   The normalized value reaches a maximum of 90% for a 2.5 μm silica particle, indicating the observed Brownian motion is approaching the definite ballistic motion that maintains a constant velocity.   In the ballistic regime, Brownian motion preserves its directionality and magnitude of the momentum after the particle receives the impulse.

The velocity of a Brownian particle is undefined if its motion is observed at the time scale of random motion since in that case it is a non-smooth curve and therefore non-differentiable.   However, this is not a limitation for ballistic Brownian motion. Figure 4 shows the velocity autocorrelation function of a 2 μm resin particle in water. At the short time scale, the velocity autocorrelation function initially decays approximately exponentially, followed by a power law decay regime.   Negative correlations are present in the regime between 10 μs to 300 μs.   This negative correlation is due to the harmonic potential since the particle has to change the direction under the restoring force when it moves close to the boundary of the trap.   The data has been fitted with the theory of Clercx and Schram (16), which describes a Brownian particle moving in a harmonic potential taking into account the fluid inertia effect (see Supporting online materials). The velocity autocorrelation function gives a statistical average of the calculated velocity. With current temporal and spatial resolutions, averaging has to be performed because of the existence of high frequency electronic noise and laser noise.   Further reduction of this noise will enable a calculation of velocity time traces directly from position time traces.   This will allow fundamental studies on a single Brownian particle, for example, verifying the Maxwell velocity distribution for a macroscopic particle in a heat bath, confirming the equipartition theorem for a Brownian particle that has been assumed in theoretical derivations but has never been directly observed experimentally.

**Conclusion**

We have observed ballistic Brownian motion for single particles with different sizes and densities and the particle's inertia effect on Brownian motion has been clearly shown. The current theory of Brownian motion has been verified down to the time scale smaller than $\tau_p$, more than two orders of magnitudes shorter than previous experiments.   For the first time, the velocity autocorrelation function of single Brownian particles is shown and we find excellent agreement with theory.   The development of an instrument that combines very high spatial and temporal resolution will also have important impact on other research fields, e.g. thermal noise imaging (17) and studies of molecular motors with the photonic force microscope (18).   Since the hydrodynamic radius of the particle can be found by fitting the data, it is feasible to observe particle growth, for example, protein adhesion to the particle.   The development also increases the probing frequency range of microrheology, which allows for example to study the high frequency relaxation of polymer networks (19).

**Acknowledgment**:   The authors like to thank Mark Raizen and Isaac Chavez for helpful discussion and the detector system.

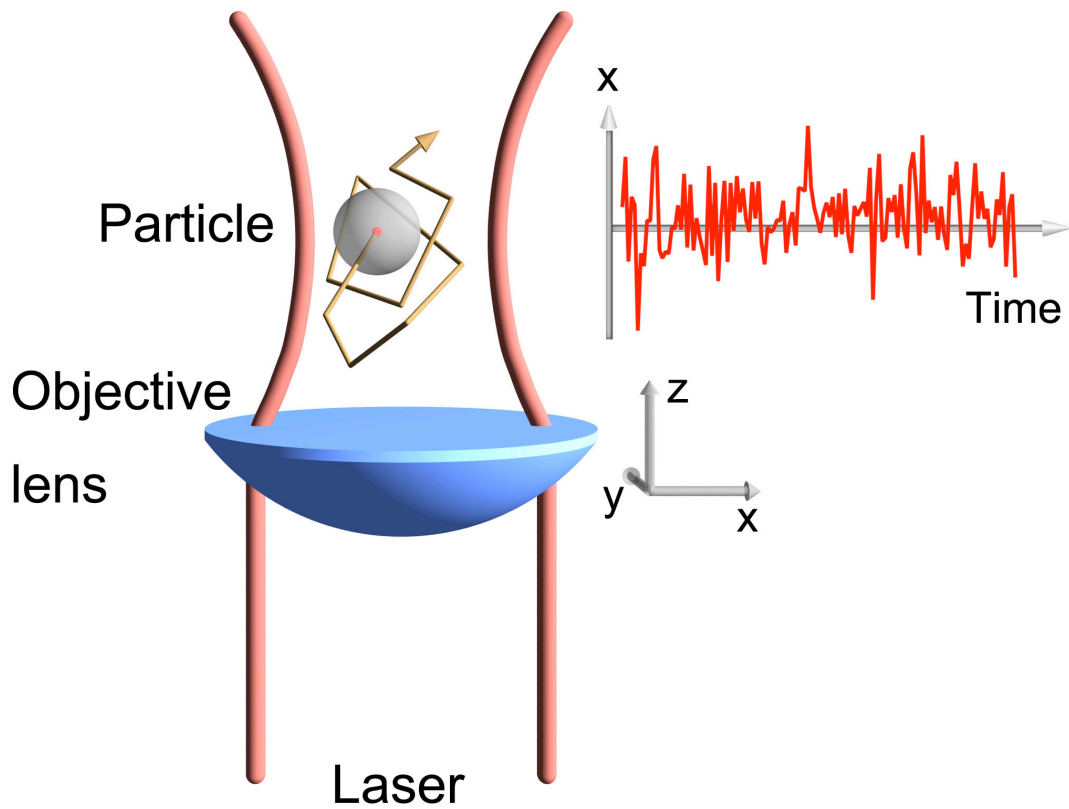

Fig. 1: Schematic of the experiment. A single micrometer-size particle is confined by an optical trap to the observation volume. While the particle undergoes Brownian motion within the trap, its position time trace along one lateral dimension is recorded with a 75 MHz bandwidth position detector.

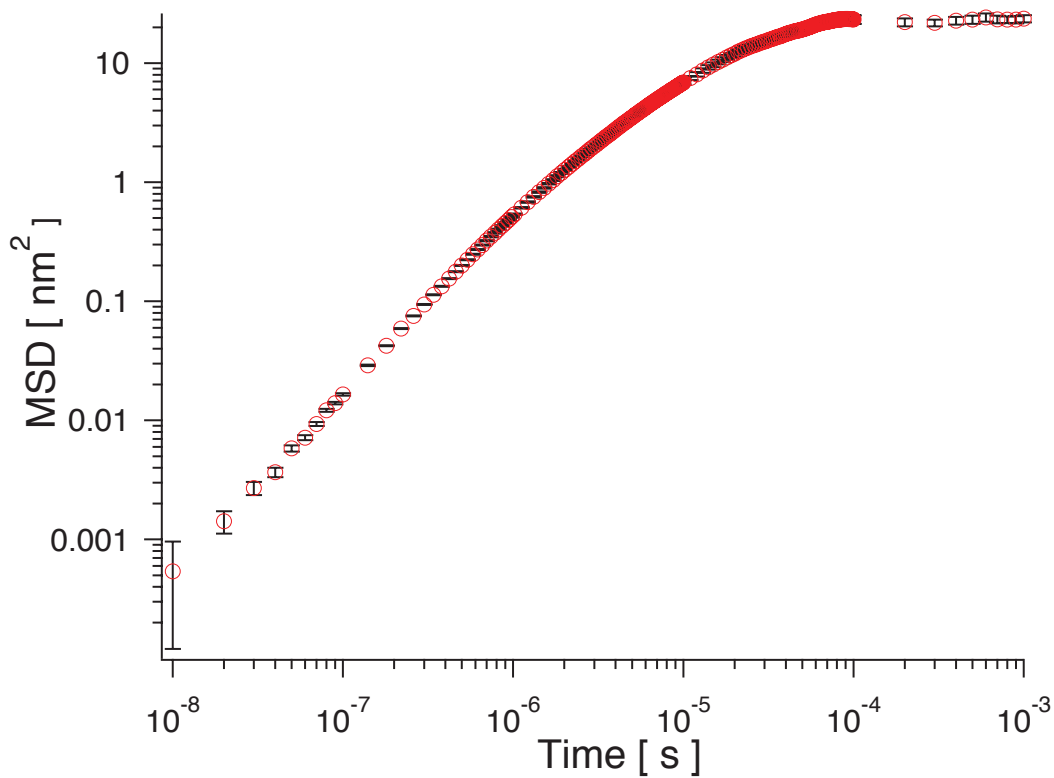

Fig. 2: MSD of a silica particle with 1 μm in diameter. Data was sampled at 100 MHz for 40 ms. The noise floor on the MSD has been subtracted and the MSD has been calibrated by fitting to Hinch theory (see Supporting online materials). Errors were calculated with a method treating data points with non-vanishing correlation (20).

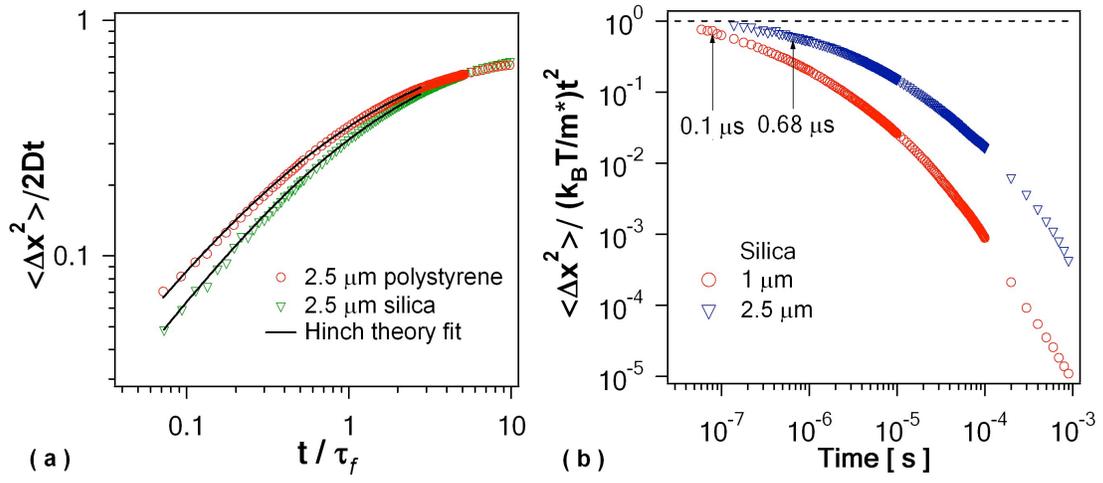

Fig. 3: (a) Normalized MSD for 2.5 μm silica (1.96 g/ml) and polystyrene (1.05 g/ml) particles. The time scale is normalized to $\tau_f$, and the MSD is normalized to the value at free diffusion regime $2Dt$. (b) MSD of 1 μm and 2.5 μm silica particles normalized to the value at the ballistic regime $t^2 k_B T/m^*$. Arrows indicate the value of $\tau_p$ for each particle.

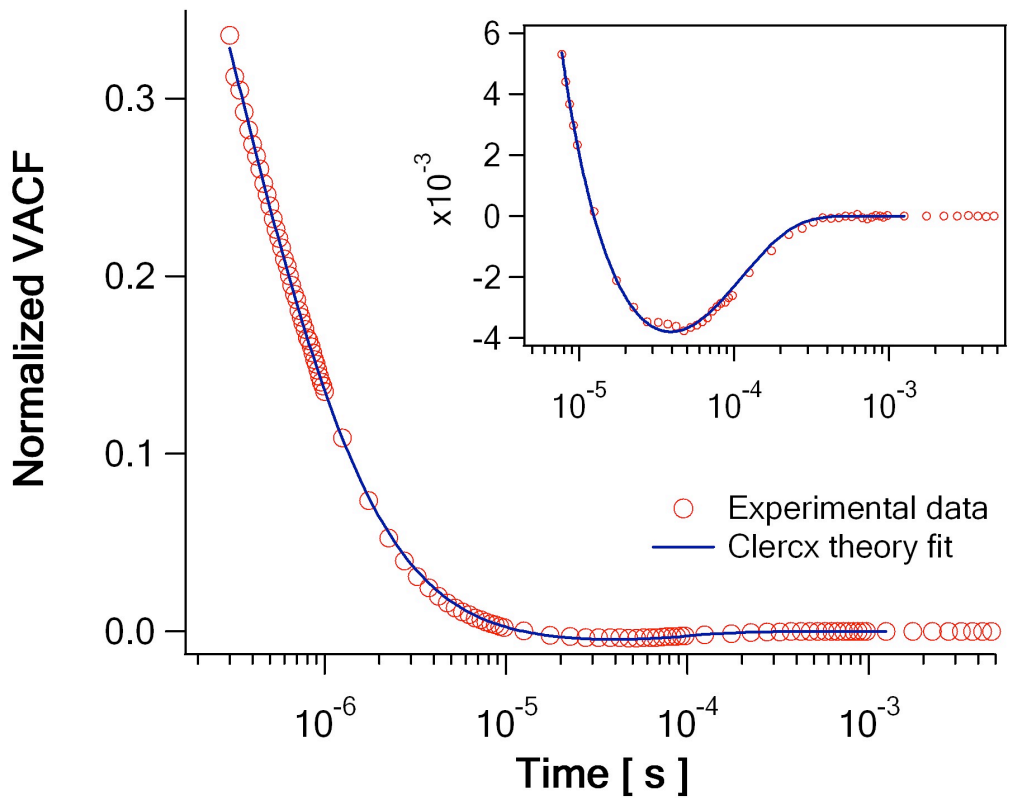

Fig. 4. Velocity autocorrelation function normalized to $k_BT/m^*$ of a 2 μm resin particle in an optical trap with spring constant 280 pN/μm. Data was averaged over 100 segments sampled at 50 MHz for 80 ms. The data above 1 μs has been blocked to 20 points per decade for fitting and display. The solid line shows Clercx theory fit to the whole experimental velocity autocorrelation function. The fit includes a calibration factor as the only fitting parameter. The stiffness of the trap is obtained from a 10 seconds of time trace sampled at 50 kHz.

**Supporting online material**

**Sample preparation.** Particles are purchased from vendors in a highly concentrated solution. We dilute the solution with deionized water to a concentration of about 1 particle per 200 x 200 x 200 $\mu m^3$ volume. The diluted solution is then injected into a circular sample chamber with a thickness of 0.8 mm. The sample chamber is formed by adhering two coverslips 15 mm in diameter to a stainless steel disk with a 11 mm center hole. The sample chamber is then attached to a three-axis scanning stage for experiments.

**Experimental setup.** The experimental setup includes an optical trap and a high bandwidth detector. A low noise laser beam is expanded and then focused by a water immersion lens into the sample chamber. Particles with higher refractive index than water will be trapped at the focal spot. Due to the weak confinement of the optical trap, the particle's position in the trap still fluctuates on the order of tens to hundreds of nanometers. The forward-scattered light and unscattered light are collected by a condenser lens. A 25 mm lens is positioned after the condenser lens to collimate the beam. The collimated beam enters the front end of the fiber bundle and is vertically divided into half. The two beams are directed to two separate photodiode on the detector by the additional optics. The intensity difference recorded at those two photodiodes can be output with a bandwidth up to 75 MHz. The difference is directly proportional to the particle's displacement in the optical trap along one lateral dimension

(14).

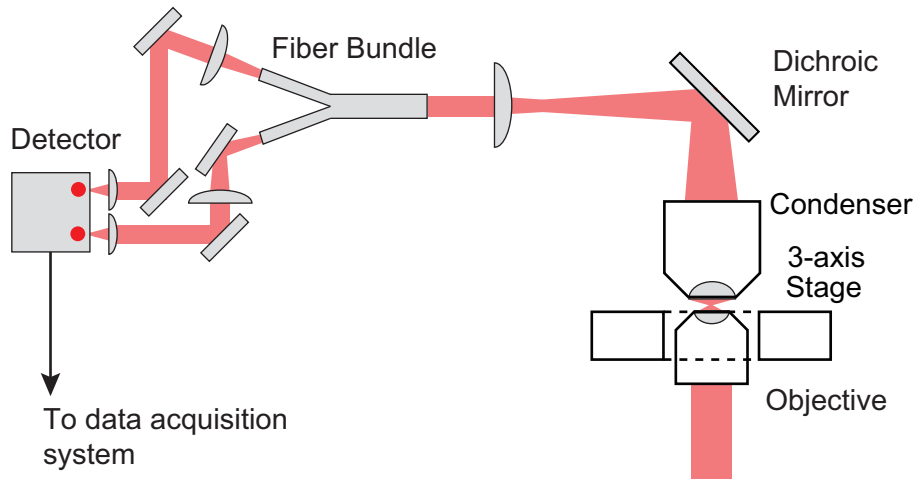

Fig. 1. Optical tweezers setup with high-bandwidth detector. The trap is formed by focusing an expanded Gaussian laser beam with a high numerical aperture objective lens. Forward scattered and unscattered light are collected by a condenser, and collimated by the first lens after a dichroic mirror. The fiber bundle splits the collimated light beam into two halves as done by a quadrant photodiode in previous laser interferometry experiments (13, 15). The differential output from the two single detectors is proportional to the trapped particle's lateral displacement. The detector system has a 75MHz bandwidth and a gain of $2 \times 10^5$.

**Noise background subtraction in MSD.** The time series in the cases with and without a particle in the trap are recorded and the MSDs are calculated for each time series. The MSD without a particle is the noise MSD that contains the contribution from all the background noise, e.g. laser, detector amplifier and digitizer. Typically, the noise is uncorrelated with the motion of the Brownian particle. So this noise MSD can be subtracted from the MSD obtained with a particle in the trap to give the true MSD of the particle.

**Fitting MSD with Hinch theory.** The data in the free diffusion region is fitted with Hinch theory. The MSD given by Hinch is the following:

$$\langle \Delta x^2 \rangle = 2Dt[1 - 2\sqrt{\frac{1}{\pi}\frac{\tau_f}{t}} + \frac{8}{9}\frac{\tau_f}{t} - \frac{\tau_p}{t} + \frac{3}{t(5\tau_f - 36\tau_p)^{1/2}}$$
$$\times (\frac{1}{\alpha_+^3} e^{\alpha_+^2 t} erfc(\alpha_+ \sqrt{t}) - \frac{1}{\alpha_+^3} e^{\alpha_-^2 t} erfc(\alpha_- \sqrt{t}))]$$

Where

$$\alpha_\pm = \frac{3}{2}\frac{3 \pm (5 - 36\tau_p/\tau_f)^{1/2}}{\sqrt{\tau_f}(1 + 9\tau_p/\tau_f)}$$

$$\tau_p = \frac{m}{\gamma}$$

$$\tau_f = \frac{\rho_f r^2}{\eta}$$

We use two fitting parameters, particle radius and a calibration factor, to convert Volts to nanometers. Weighted least square with standard deviation is chosen as the fitting method, greatly increasing the stability of the fit. Measurements have been done on silica, polystyrene and resin particles with various sizes. The Hinch theory fits the data in the free diffusion region very well for all the particles we have tested.  The particle radius from the fit falls in the size range given by the manufacturer for silica particles. For polystyrene and resin particles, the average fitted particle radius is larger than the specified mean particle size.  This is possibly because the polymer particles swell in water with water molecules penetrating into the particle surface and thus increasing its effective size.  The fitting with these two parameters is the best method to find their true values, since the particle radius and the calibration factor are closely related.  In other optical tweezer calibration methods, one takes the particle radius as the mean radius given by the manufacturer, however the size of the particles typically varies by 10% to 15%.

**Temporal and spatial resolution for different particles.**  The temporal resolution is defined as the time at which the signal-to-noise ratio reaches unity on the MSD curve. The spatial resolution inferred from the MSD is obtained by taking the square root of that value.  This resolution is not the same as the resolution in electron or light microscopy that one can resolve individual atoms or cells. Neither is it the same as the precision in single particle tracking quantifies how precise one can track the center position of an individual particle. Here, resolution means the average displacement of the particle can be ascertained to this value with our technique. The last few points on MSD at the long time scale have large errors due to low statistics in calculating the MSD, which is a result of limited sampling time.

**Calculation of velocity autocorrelation function.** The time series of the position traces are sampled at 50 MHz in the resin particle experiment. However, the temporal resolution on resin particles is only 5 MHz. Therefore, high frequency noise is present in our measurement. A boxcar smooth with size of 10 points is applied to the data to remove the high frequency noise above 5 MHz. The velocity autocorrelation function is calculated on this smoothed data trace and an average velocity autocorrelation function is then obtained from 100 data traces sampled consecutively from the same particle. The length of a single data trace is limited to the onboard memory on our digitizer which is 4 million points in a single shot.

**Calibrating the stiffness of the optical trap in resin experiment.** We first find the particle radius by fitting the velocity autocorrelation function at the free diffusion region with Hinch theory. The velocity autocorrelation function given in the Hinch theory is as follows:

$$\phi(t) = \frac{k_B T}{m^*(\alpha_+ - \alpha_-)}[\alpha_+ \exp(\alpha_+^2 t) erfc(\alpha_+ \sqrt{t}) - \alpha_- \exp(\alpha_-^2 t) erfc(\alpha_- \sqrt{t})]$$

Where m* is the effective mass of the particle given as:

$$m^* = m + \frac{1}{2}m_0 = \frac{4}{3}\pi r^3 \rho_p + \frac{2}{3}\pi r^3 \rho_f$$

The fitting determines the particle radius and the calibration factor. Additionally, a time trace is sampled at 50 KHz for 10 seconds so that the particle is allowed to fully explore the trapping potential. Calibration on the low frequency data using the particle radius from the previous fit is carried out to find the stiffness of trap. The particle radius is 1.202 μm and the stiffness of the trap is 2.80 x $10^{-4}$ N/m.

**Fitting to the velocity autocorrelation function with Clercx and Schram theory.**

Clercx and Schram studied Brownian motion in a harmonic potential taking into account the effect of fluid inertia. They replaced the steady state Stokes' friction with a time-dependent friction force as a result of the co-moving fluid and its acceleration history. The equation of motion is given as follows:

$$m\ddot{x}(t) = -6\pi\eta r\dot{x}(t) - \tfrac{1}{2}m_0\ddot{x}(t) - 6r^2\sqrt{\pi\rho\eta}\int_0^t \frac{1}{\sqrt{t-\tau}}\ddot{x}(t)d\tau$$
$$-Kx(t) + R(t)$$

The velocity autocorrelation function obtained by Clercx and Schram is given as follows:

$$\phi(t) = \frac{k_B T}{m^*}\left[\frac{A^3 \exp(A^2 t)\,\mathrm{erfc}(A\sqrt{t})}{(A-B)(A-C)(A-D)} + \frac{B^3 \exp(B^2 t)\,\mathrm{erfc}(B\sqrt{t})}{(B-A)(B-C)(B-D)}\right.$$
$$\left.+ \frac{C^3 \exp(C^2 t)\,\mathrm{erfc}(C\sqrt{t})}{(C-A)(C-B)(C-D)} + \frac{D^3 \exp(D^2 t)\,\mathrm{erfc}(D\sqrt{t})}{(D-A)(D-B)(D-C)}\right]$$

Where $A$, $B$, $C$ and $D$ are the four roots of the equation

$$(\tau_p + \tfrac{1}{9}\tau_f)x^4 - \sqrt{\tau_f}\,x^3 + x^2 + 1/\tau_k = 0$$

$\tau_k = \gamma/K$ is the autocorrelation time of the trap and $K$ is the stiffness of the harmonic

potential. The whole data is fitted with the above velocity autocorrelation function with a calibration factor. The fitted scaling factor agrees with the one obtained from the fitting with Hinch theory.